# Deducing the Composition of Venus Cloud Particles with the Autofluorescence Nephelometer (AFN)

**Darrel Baumgardner** [1,2,*], **Ted Fisher** [2], **Roy Newton** [2], **Chris Roden** [2], **Pat Zmarzly** [2], **Sara Seager** [3,4,5], **Janusz J. Petkowski** [3], **Christopher E. Carr** [6], **Jan Špaček** [7], **Steven A. Benner** [7], **Margaret A. Tolbert** [8], **Kevin Jansen** [8], **David H. Grinspoon** [9] **and Christophe Mandy** [10]

[1] Droplet Measurement Technologies, LLC, 2400 Trade Centre Ave, Longmont, CO 80503, USA
[2] Cloud Measurement Solutions, LLC, 415 Kit Carson Rd, Unit 7, Taos, NM 87571, USA; tempusfinite@gmail.com (T.F.); window_shade@yahoo.com (R.N.); chris.roden@gmail.com (C.R.); pzmarzly@comcast.net (P.Z.)
[3] Department of Earth, Atmospheric and Planetary Sciences, Massachusetts Institute of Technology, 77 Massachusetts Avenue, Cambridge, MA 02139, USA; seager@mit.edu (S.S.); jjpetkow@mit.edu (J.J.P.)
[4] Department of Physics, Massachusetts Institute of Technology, 77 Massachusetts Avenue, Cambridge, MA 02139, USA
[5] Department of Aeronautics and Astronautics, Massachusetts Institute of Technology, 77 Massachusetts Avenue, Cambridge, MA 02139, USA
[6] School of Aerospace Engineering and School of Earth and Atmospheric Science, Georgia Institute of Technology, Atlanta, GA 30312, USA; cecarr@gatech.edu
[7] Firebird Biomolecular Sciences, LLC, 13709 Progress Blvd N134, Alachua, FL 32615, USA; jspacek@firebirdbio.com (J.Š.); sbenner@ffame.org (S.A.B.)
[8] Department of Chemistry and CIRES, University of Colorado-Boulder, 216 UCB, Boulder, CO 80309-0215, USA; tolbert@colorado.edu (M.A.T.); kevin.jansen@colorado.edu (K.J.)
[9] Planetary Science Institute, 1700 East Fort Lowell, Suite 106, Tucson, AZ 85719, USA
[10] Rocket Lab, 3881 McGowen Street, Long Beach, CA 90808, USA
* Correspondence: darrel.baumgardner@gmail.com

**Abstract:** The composition, sizes and shapes of particles in the clouds of Venus have previously been studied with a variety of in situ and remote sensor measurements. A number of major questions remain unresolved, however, motivating the development of an exploratory mission that will drop a small probe, instrumented with a single-particle autofluorescence nephelometer (AFN), into Venus's atmosphere. The AFN is specifically designed to address uncertainties associated with the asphericity and complex refractive indices of cloud particles. The AFN projects a collimated, focused, linearly polarized, 440 nm wavelength laser beam through a window of the capsule into the airstream and measures the polarized components of some of the light that is scattered by individual particles that pass through the laser beam. The AFN also measures fluorescence from those particles that contain material that fluoresce when excited at a wavelength of 440 nm and emit at 470–520 nm. Fluorescence is expected from some organic molecules if present in the particles. AFN measurements during probe passage through the Venus clouds are intended to provide constraints on particle number concentration, size, shape, and composition. Hypothesized organics, if present in Venus aerosols, may be detected by the AFN as a precursor to precise identification via future missions. The AFN has been chosen as the primary science instrument for the upcoming Rocket Lab mission to Venus, to search for organic molecules in the cloud particles and constrain the particle composition.

## 1. Introduction

The Venera 9, 10 and 11 landers and Pioneer missions to Venus in the 1970s, and the Venera and Vega missions in the 1980s have produced a wealth of information about the particles in the atmosphere of Venus, including size distributions and estimates of refractive index along with speculations about composition [1–13]). During the Pioneer mission direct measurements of the equivalent optical diameter (EOD) were made with an instrument, designated the Large Cloud Probe Spectrometer (LCPS) [1] that combined a single particle light scattering spectrometer with an optical array probe (OAP). The size distributions from the LCPS were constructed by binning the single particle data into fixed size intervals; hence, the information from individual particles was lost. Likewise, on the Venera, Vega and Pioneer missions, the optical properties were deduced from measurements with multiwavelength, backscattering nephelometers that were fit to models in order to derive refractive index. An additional speculation was that some fraction of the cloud particles were not liquid but rather crystalline in shape. This deduction was made based both on data from the nephelometers and from a mismatch in sizes between the size ranges in the LCPS.

Whereas the analyses that were applied to the single particle and ensemble nephelometers were quite clever, and maximized the information available, the conclusions about particle refractive index and shape remain open to question due to the measurement limitations and uncertainties of the LCPS and backscatter nephelometers, and the assumptions that were necessary in order to reach these conclusions. Of particular interest was the range in estimated refractive indices that have evolved from these analyses. The refractive index estimates range from as small as 1.32 in the lower cloud regions to as large as 1.50 in the haze found just below the lowest cloud layers. Other, non-in situ studies, which used reflected, polarized sunlight from the Venus atmosphere [14–16], derived a refractive index of 1.46 ± 0.015 at a wavelength of 365 nm, consistent with a cloud droplet composition of sulfuric acid. Hence, particles with refractive indices that deviate from that of $H_2SO_4$ would likely have a different particle composition, perhaps $H_2SO_4$ mixed with some other atmospheric constituent.

Many reasons have been hypothesized [17] for the departure in values of the refractive indices that were derived from what had been previously reported and two explanations for this difference have been proposed: (1) some of the particles could be light-absorbing at the UV wavelength, i.e., their refractive indices would be complex or (2) the lower refractive index particles might not be spherical, perhaps crystalline in shape. Either of these explanations would partially explain why the assumptions of sphericity or of refractive indices with no imaginary component would lead to unexpectedly lower refractive indices.

In addition to uncertainty about the properties of the Venus cloud particles, other open questions involve the possible presence of organic compounds in the atmosphere of Venus [18] that could serve as possible precursors for exotic life forms. This has led to the development of an exploratory mission, developed by Rocket Lab, that will drop a small probe, instrumented with a single-particle optical spectrometer (here the term "spectrometer" refers to separation by size), into Venus's atmosphere to address some of these questions [19].

Similar to the LCPS that was deployed for the Pioneer Venus mission, a sensor has been developed to make single-particle measurements that will characterize some of the physical, optical and chemical properties of Venus cloud droplets. The design criteria for the Autofluorescence Nephelometer (AFN) were driven primarily by the results from the Pioneer mission and aim to address some of the questions that were raised by its measurements. In particular, the AFN will achieve the following:

- Detection and derivation of the equivalent optical diameter (EOD) of individual particles from 1–10 μm with an accuracy of ±20% (limited by uncertainties in shape and refractive index).

- Derivation of particle number concentration < 5000 cm$^{-3}$ over the specified size range with an accuracy of ±30% (limited by uncertainties in particle velocity).
- Estimation of complex refractive index (CRI) with an accuracy of ±30%.
- Quantification of particle asphericity with an accuracy of ±30% (limited by uncertainties in orientation and shape of non-spherical particles).
- Direct measurement of fluorescence from individual particles at the wavelength band between 470–520 nm with an accuracy of ±10% (limited by background light, mass of organic material and fluorescence efficiency at the excitation wavelength of 439–441 nm).

## 2. Materials and Methods

Figure 1a is a cartoon that illustrates the basic concept of how the AFN detects individual particles and extracts information about their properties. As illustrated in Figure 1b, the AFN projects a collimated, focused, linearly polarized, 439–441 nm laser beam through a fused silicate window that is installed on the outer surface of the space capsule. The center of focus (COF) of the laser beam is 20 cm from the collection optics that are located inside the capsule, protected by a pressure vessel that maintains the instrument at constant pressure and temperature range from approximately 0–35 °C. The sample volume of the AFN was designed to intercept particles flowing past the capsule window, outside the boundary layer.

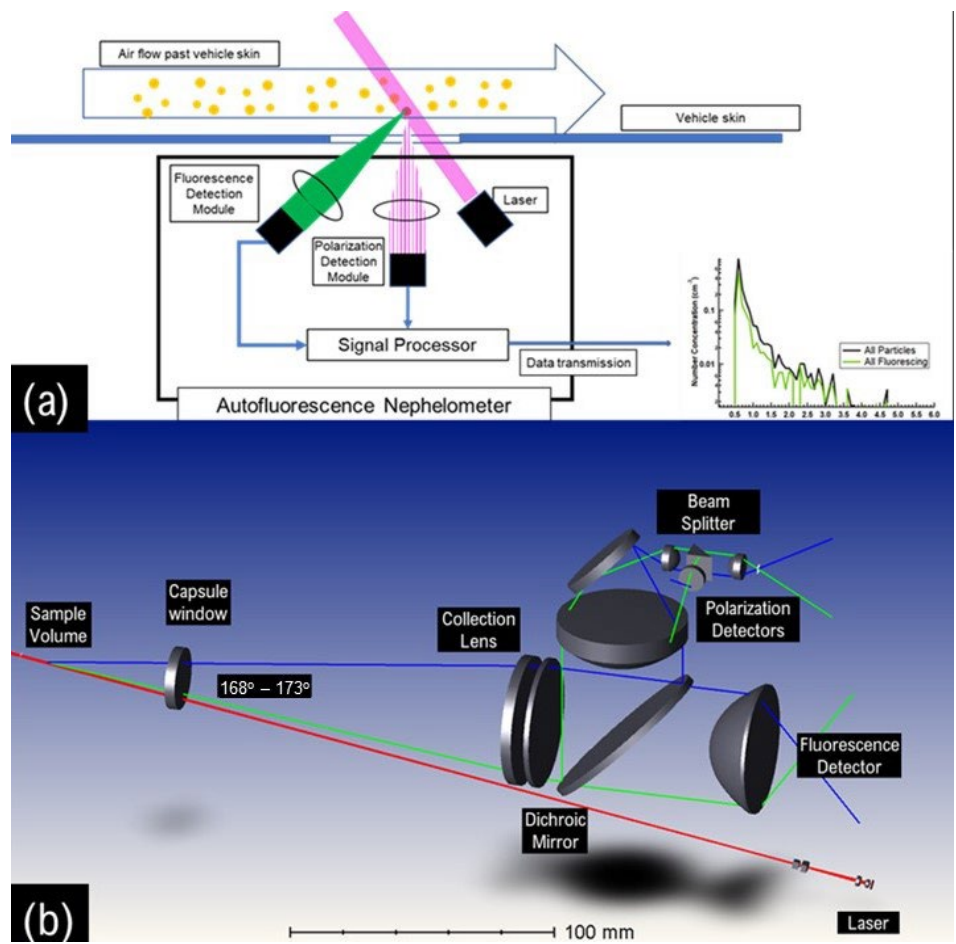

**Figure 1.** (**a**) This cartoon illustrates the basic features of the AFN in which a 440 nm diode laser projects a focused beam out a window into the air-stream where it is intercepted by individual cloud droplets that scatter light. Some may also fluoresce. The scattered and fluorescent light is collected and processed to derive size distributions and complex refractive indices, (**b**) This diagram shows the optical components of the AFN.

Individual particles that pass through the laser beam scatter light in all directions and the intensity of the scattered light varies with respect to the angle. The angular sensitivity depends on the incident laser's wavelength and the properties of the illuminated particle, i.e., its size, shape and complex refractive index (CRI). The CRI is determined by the composition of the particle. The AFN collects the photons that are backscattered (168–173°) through the same window through which the laser is projected and focuses them onto a dichroic mirror that directs them to a spatial filter followed by a beam splitter. The spatial filter restricts the volume of air that is seen by the collection optics, thereby defining the optical volume that is used to calculate the number concentration, as well as minimizing the occurrence of two particles being coincident in the view volume and erroneously being counted and sized as one particle.

The beam splitter separates the scattered light into two, approximately equal components, one that is transmitted through a filter that allows only light with the same plane of polarization as the incident laser beam and the other through a filter that passes only light with polarization perpendicular to the incident light. The two, scattered light components impinge upon avalanche photodetectors (APD) that convert the photons to an electrical current. These two polarized light components are designated "Ppol" and "Spol", respectively.

In addition to scattering light, some particles, depending on their composition, will also autofluoresce when excited at the 440 nm wavelength. As with the scattered light, the fluorescence is omnidirectional, and some of the fluorescence photons are transmitted back through the capsule window where they are directed towards the dichroic mirror that is designed to pass, rather than reflect, those photons with a wavelength of 470–520 nm. The fluorescence light is directed onto a photomultiplier tube (PMT) that converts the photons to an electrical signal. The excitation and emission wavelengths chosen for the AFN were motivated by laboratory work performed by Firebird Biomolecular Sciences (FBS) which demonstrated that a wide variety of simple organic compounds (OC), when reacted with concentrated sulfuric acid, will give a mixture of more complex products that are fluorescent when excited over a large range of UV and visible excitation wavelengths. The FBS team generated various mixtures of sulfuric acid (>70%) with different OCs. They found that virtually all the tested OCs (including single carbon species such as formaldehyde) were converted into mixtures with coloration and fluorescence spectra profiles highly dependent on the reaction conditions. The FBS team analyzed these samples using 3D fluorescence spectroscopy (excitation and emission range between 200 nm to 700 nm). The results from their studies are summarized in Figure 2 where each marker represents position of fluorescence intensity peak maxima provided by products generated from various OC reacting in sulfuric acid under various, Venusian cloud-like conditions. Peak maxima overlaps are indicated by darker circles. From this figure we see that the three excitation wavelengths that lead to the most compounds that fluoresce are ~260 nm, ~350 nm and ~450 nm. Any of the three excitation wavelengths would be good candidates for the AFN but only the 440 nm excitation wavelength was available in a diode laser compact enough to meet the AFN's weight, volume and power design criteria.

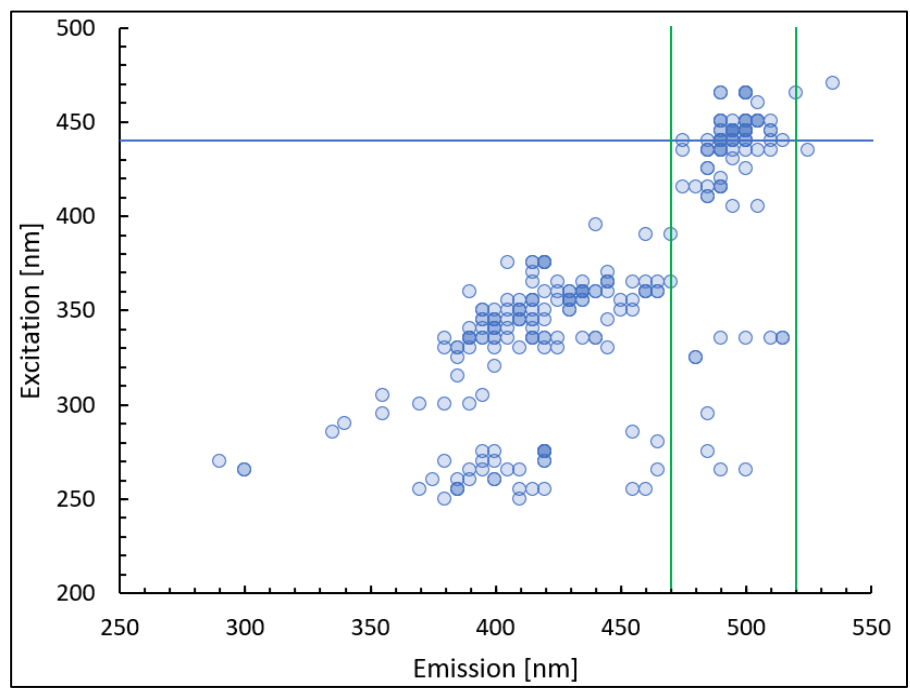

**Figure 2.** The fluorescing emission maxima wavelength as a function of excitation wavelength is shown for a variety of sulfuric acid mixtures with organic compounds. Each marker represents fluorescence intensity maximum yielded from one reaction conditions. The darkness of the marker indicates the number of instances with overlapping fluorescence peak maxima. The blue horizontal line represents the 440 nm excitation band of the blue Autofluorescence Nephelometer (AFN) laser and the green lines demarcate the range of possible wavelengths where the organic might emit fluorescence (in a cyan to green spectrum).

The electrical signals from the APDs and PMT are amplified, digitized and processed to produce five pieces of information: (1) the peak amplitude from the Ppol detector, (2) the peak amplitude from the Spol detector, (3) the peak amplitude from the fluorescence detector, (4) the duration of time that the particle was in the sample volume (transit time) and (5) the elapsed time since the previous detected particle. The size of each particle is derived from the sum of the S and P signals using Mie theory that relates the scattering cross section of a spherical particle to its geometric diameter. In order to apply this theory, we have to assume the particle is spherical. In addition, we must know the wavelength of the incident laser as well as the refractive index of the particle and the collection angle of the optical system. As explained below, the refractive index is derived from selected, individual particles and from an ensemble of particles. As shown in Figure 1b, the light is collected over a cone defined by the 20 cm distance of the COF from the optics and the 25 mm diameter of the collection lenses. The angles of this collection cone are 168–173°. Figure 3 illustrates an example of the relationship between the scattering cross section and particle diameter calculated for these angles and two real (non-absorbing) refractive indices. Note that whereas the curves in this example are for the S-polarized scattered light, similar curves are calculated for the P-polarized light; however, the relative magnitudes of the peaks and valleys, and their location with respect to one another, are different than for the S-Polarized cross sections. The differences in the S and P-pole features are what allows us to estimate the CRI with greater accuracy than previous Venus mission measurements.

In Figure 3, the red lines illustrate one of the features of Mie theory, i.e., the non-monotonic nature of the relationship between scattering cross section and size whereby some particles of different size can have the same scattering cross section. This ambiguity introduces uncertainty into the derived particle diameter and also explains why we refer to an equivalent optical diameter (EOD) when referring to the derived size.

Taking a closer look at the oscillating nature of the Mie relationship, and comparing the curves for two refractive indices, we observe that the relative locations of the peaks and valleys, and their respective maxima and minima, are different for individual refractive indices. A similar pattern can be seen in the Ppol curves (not shown here). These oscillating features are what the AFN will use to extract an estimate of the CRI because they represent a unique set of identifying structures that are associated with each real and imaginary refractive index. As the AFN samples the cloud particles it will construct a frequency distribution of peak signals from the Ppol and Spol detectors. These distributions will be input to a neural network that will be trained to associate these patterns with average CRIs.

It is important to highlight the differences between the AFN, the single particle nephelometer on the Pioneer and Vega missions, and the multiwavelength, backscattering nephelometers that have also been deployed on the Venera, Pioneer and Vega missions. Individual particles were measured on the Pioneer mission with the LCPS, which implemented light scattering and imaging, and on the Vega mission with a photoelectric sensor that also measured light scattered from single particles. Neither of these instruments, however, measured the shape of the cloud droplets in sizes smaller than 20 µm, nor could they derive CRI from the measurements in the way the AFN is designed to do. Likewise, the backscattering nephelometers provide measurements from ensembles of particles from which estimates of the size distributions and refractive indices are derived. These derived quantities require a number of assumptions about the composition, shape of the individual particles and shape of the size distributions. The solutions to the models that fit the measurements are not well-constrained and hence the estimated size distributions and refractive indices have large error bars.

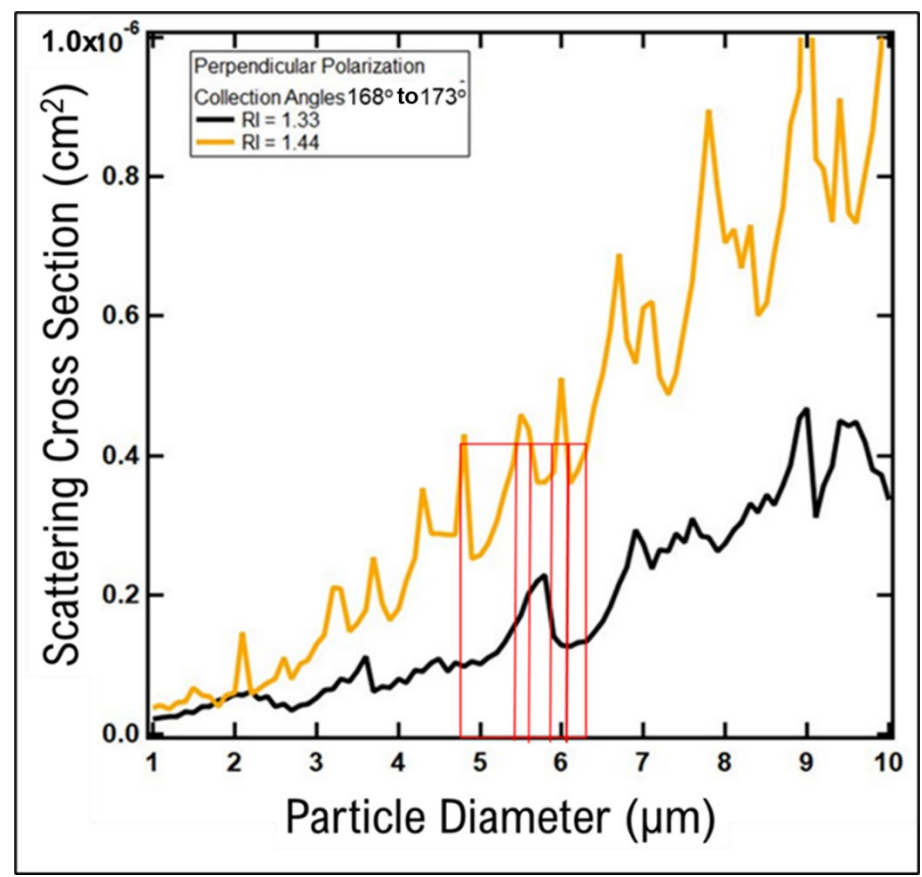

**Figure 3.** The relationship between the light scattered by an illuminated single particle, as a function of size, is non-monotonic and highly sensitive to the refractive index of the particles, as highlighted by the vertical red lines.

The strength of the AFN approach that is used to derive shape, size and refractive index is that from the individual particles a fluorescence event is compared with the simultaneous, polarized scattering signals to evaluate how the CRI extracted from the scattering signals corresponds to the fluorescence intensity. The laboratory studies have already demonstrated that some organic compounds will fluoresce and these same compounds will absorb at 440 nm, i.e., the derived refractive index should have an imaginary component corresponding to how much of the organic compound is mixed with the sulfuric acid in the cloud droplet. The ratio of the S-polarized light to the P-polarized light is analyzed for each particle to estimate the CRI. In addition, the frequency distribution of the S and P; polarization events are analyzed to extract average CRIs as well as the CRIs of individual particles. This dual-pronged analysis constrains the range of possible CRIs and greatly increases the accuracy of the derivation.

The AFN was selected for the Rocket Lab Venus mission [19], the first of a series of missions envisioned by the Venus Life Finder (VLF) team [20], because of the information it would provide and its small size. Figure 4 shows the proposed system that will carry the AFN through the Venus atmosphere. The outer capsule measures 40 cm in diameter while the inner pressure vessel where the AFN will be mounted is 22 cm. The current design specifications have the AFN weighing a little less than one kilogram with a power consumption less than 50 W. These are critical parameters because of weight and balance issues coupled with the limited amount of power that will be available from the battery system that also has to power the spacecraft's transmitter that will be sending data back to Earth.

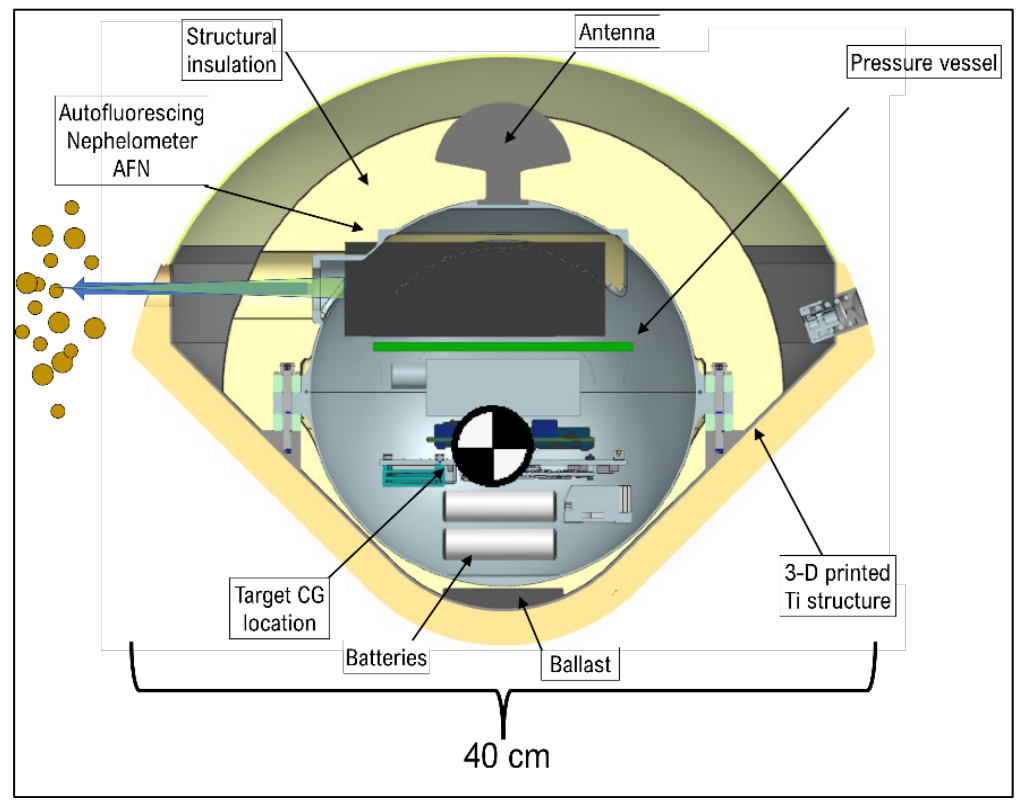

**Figure 4.** The cross section of the capsule that will carry the AFN through the Venus Atmosphere. Figure courtesy of Rocket Lab.

The other constraint that poses a major challenge is the amount of information that will be transmitted from the AFN via the spacecraft's communication system. Once the capsule enters the atmosphere, and the AFN is activated, the length of time during which the instrument and the capsule are expected to operate is approximately five minutes. As it descends the temperatures rapidly increase to the point when the electronics can no longer function. Due to the limited bandwidth and power of the transmitter, the AFN has been allocated a maximum of 125 bytes/second, or approximately 30,000 bytes of information to document the particle size distribution, CRI, and fluorescence as a function of temperature and altitude. To address this limitation, we are implementing a neural network (Figure 5) that will integrate the information from the three detectors, and the associated frequency distribution of their accumulated values, to derive the critical information to transmit. The network is being trained using laboratory measurements (see next section) and theoretical calculations.

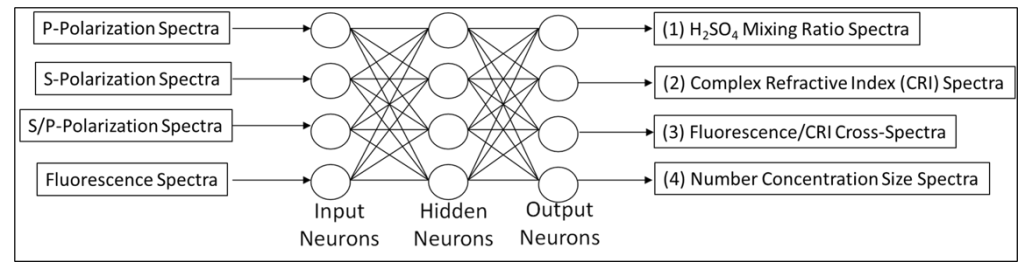

**Figure 5.** Machine learning will be implemented in the real time processor of the AFN to extract the critical information from the measurements.

In addition to the information derived from the neural network, selected housekeeping temperatures and voltages will be transmitted along with raw data that will help us to better interpret the measurements.

### 3. Results and Discussion

The development of the AFN was a three-stage process: (1) Optical, electrical and mechanical design of the instrument to meet the proposed deliverables, (2) Fabrication of a proof-of-concept, laboratory prototype (LAFN) and (3) Fabrication and environmental testing of the mission-ready AFN (MAFN). The LAFN, shown in Figure 6, is currently being tested in a laboratory at the University of Colorado. Thus far, several weeks of testing have challenged the LAFN with sulfuric acid droplets generated with an atomizer that produces a polydispersed cloud of droplets. Solutions were prepared composed of 70%, 75%, 80%, 85%, 90% and 95% $H_2SO_4$. Each of these solutions were further divided into mixtures with 0, 0.1, 1 and 10 mg ml$^{-1}$ of formaldehyde ($CH_2O$), that reacts in concentrated sulfuric acid to yield complex fluorescent organics. The droplets were directed down a flow tube where they passed through the optical sample volume of the LAFN. The signals from the detectors were digitized and sent to the signal processing electronics that encoded and then transmitted them to the data system that displayed droplets' size distributions in real time and recorded the data for subsequent processing and analysis.

Figure 7 illustrates two representative frequency distributions of the light scattering intensities detected by the S and the P-polarization detectors. We display the measurements this way to be similar to what is shown in Figure 3, i.e., highlighting the peaks and valleys that are due to the natural fluctuations and the Mie ambiguities. The top and bottom panels are measurements of pure 70% and 95% $H_2SO_4$, with no $CH_2O$ added.

There are a number of features to note in Figure 7. Three curves are drawn in each panel: (1) Spol scattering (black), (2) Ppol scattering (blue) and (3) the sum of the S and P signals (green). The Spol signals are smaller than the Ppol because spherical particles scatter the majority of their light in the same plane as the incident light. When the Spol signal intensities approach those of the Ppol, this indicates that the particles are no longer spherical. The yellow and orange boxes draw the reader's attention to peaks and valleys in both Spol and Ppol intensity distributions. These features shift with the composition in the Spol, Ppol and S + Ppol, but the patterns are unique to each composition. This is because the 70% solution of $H_2SO_4$ has a refractive index closer to water (1.33-0.0i) while the 95% $H_2SO_4$ is closer to pure $H_2SO_4$ (1.46) at the 440 nm wavelength [21]. These multiple features in the scattered light intensity are the unique patterns by which we extract the CRI and since the CRI is directly related to the composition of the droplets [21], we will be able to estimate the CRI and composition with the Neural Network in real time.

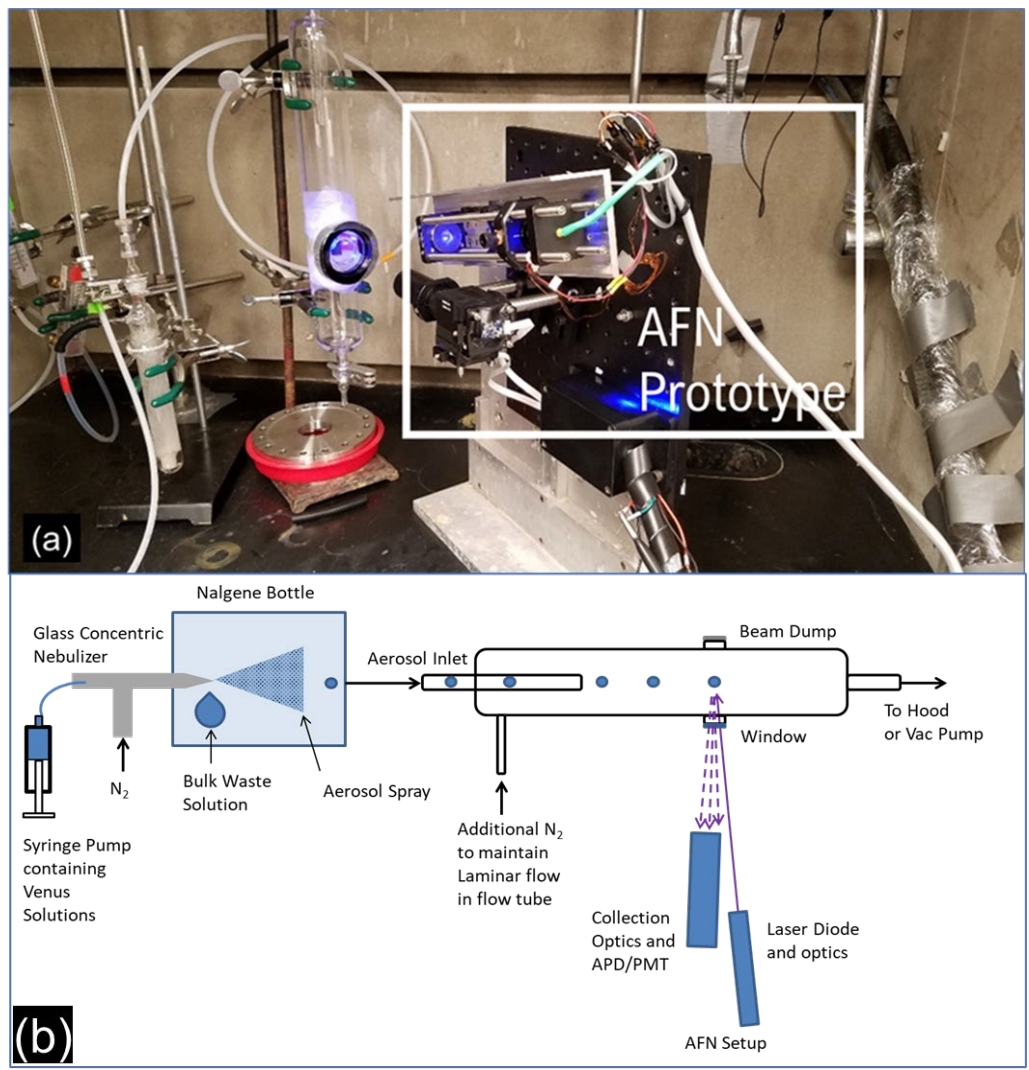

**Figure 6.** (**a**) The laboratory prototype of the AFN (LAFN) was evaluated in the laboratory where it was tested (**b**) with a polydispersed spray of droplets composed of mixtures containing $H_2SO_4$ and $CH_2O$. The collection optics direct the light to two avalanche photo detectors (APD) for the scattered, polarized light and to a photomultiplier tube (PMT) for fluorescence emissions.

Not shown here, and a subject of a separate work, are additional studies of the autofluorescence of the sulfuric acid mixtures measured in the lab. Unlike the light scattering whose intensity is proportional to the droplet size as well as its CRI, fluorescence does not necessarily scale with the size of the droplets or concentration of organic material. The fluorescence will be a complementary signal that will be used to identify the possible presence of organic molecules in individual droplets to compare with the CRI that will also indicate the presence of OC. The reason is that pure $H_2SO_4$ does not fluoresce at 440 nm, and also will not absorb at this wavelength, hence the derived CRI will have no imaginary component. If the derived CRI does have an imaginary component, this will indicate that something in the droplet is absorbing at the 440 nm wavelength.

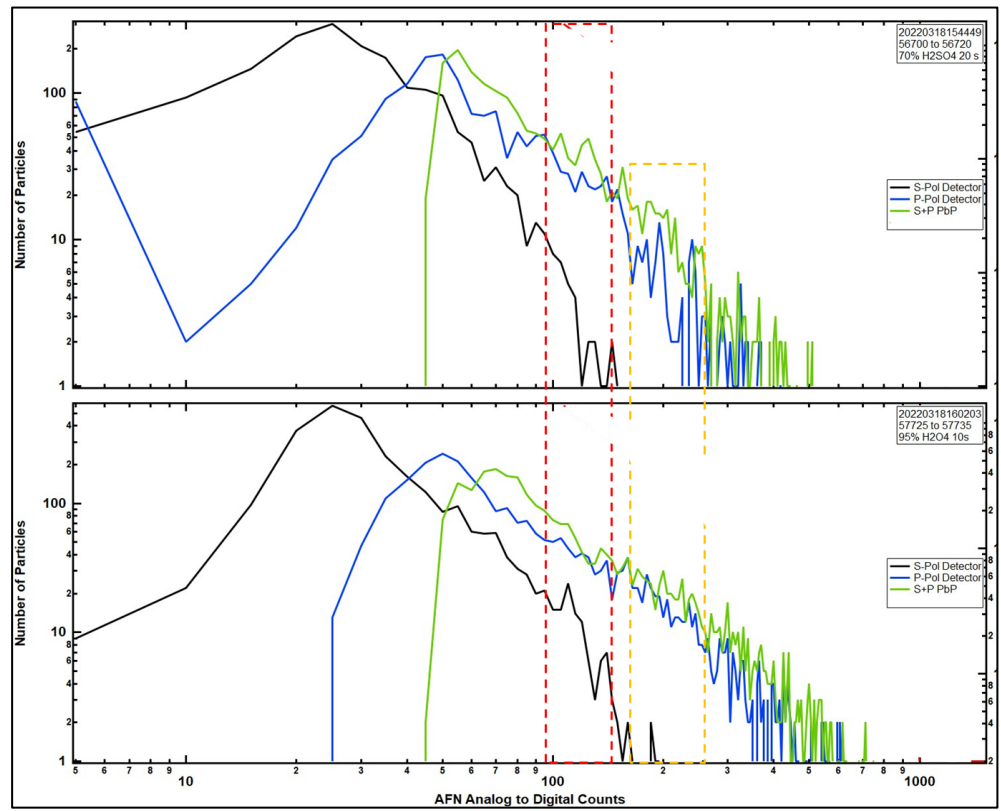

**Figure 7.** The number of particles, detected by the AFN, drawn as a function of the polarized light scattering intensity, is shown to illustrate how the Mie oscillations can be observed in these frequency distributions. The red and orange boxes highlight where the peaks and valleys have shifted due to difference in the mixing ratios.

The LAFN is currently being evaluated with more reference material, i.e., sulfuric acid mixtures with OC, which will be used for training the neural network. The MAFN has been assembled in the cleanroom facilities and is undergoing environmental testing before being delivered to the Rocket Lab facilities (Figure 8). The MAFN will be integrated with the spacecraft pressure vessel and undergo further environmental tests at the Rocket Lab facilities. At the time of this writing, the MAFN is expected to be launched in May, 2023, reaching the atmosphere of Venus five months later. The capsule will be released into the Venus atmosphere on the night-time side in order to minimize any background light at the wavelength of MAFN measurements.

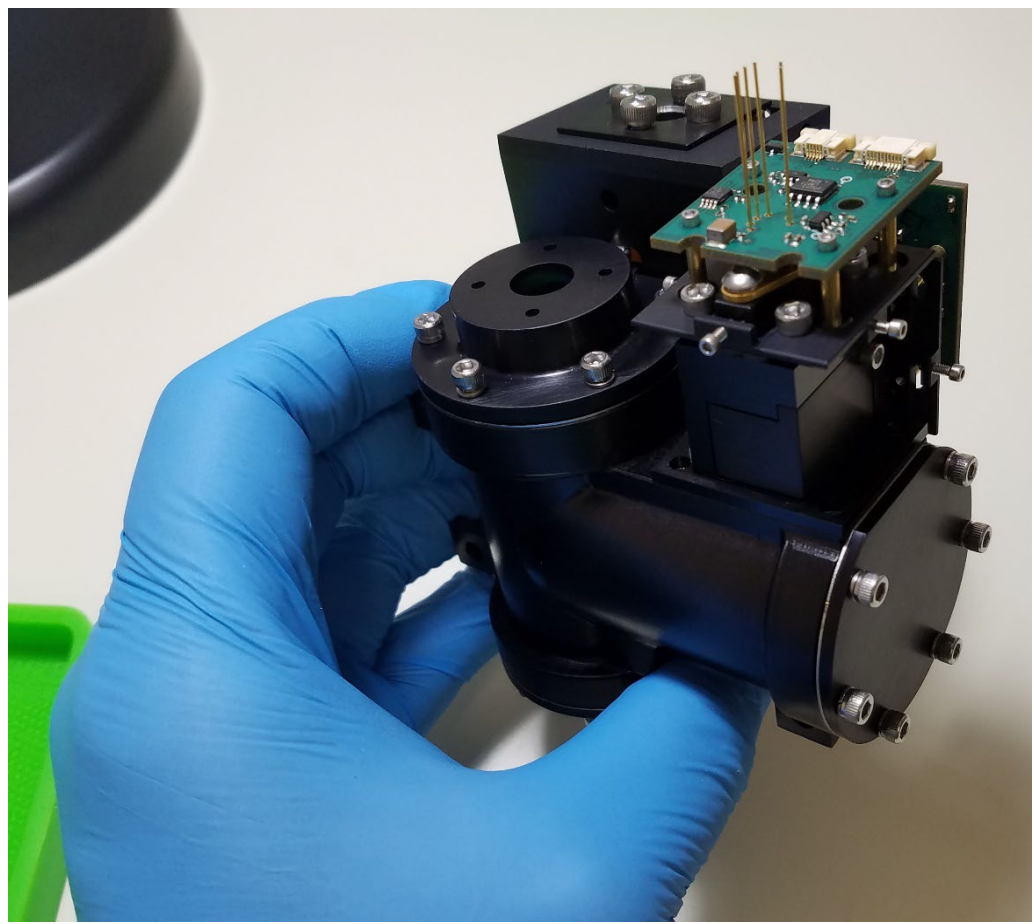

**Figure 8.** In this photo, the AFN is at 80% completed assembly. Here, is shown the entire optical assembly with detector electronics.

## 4. Summary

The single-particle Autofluorescence Nephelometer (AFN) is the sole instrument on the Rocket Lab Mission to Venus [19], with a target launch date of May 2023. The AFN is designed to induce and detect autofluoresence of organic molecules in the Venus cloud particles, by using an excitation wavelength of 440 nm and a detector at 470–520 nm. The AFN is also designed to measure the backscattered polarized radiation in the angular range of 168-173 degrees in order to provide constraints on particle number, size, shape, and composition. The AFN is the first instrument to directly examine cloud particles on Venus in nearly four decades.

**Author Contributions:** Conceptualization, D.B.; methodology, D.B., T.F., R.N., C.R. and P.Z.; software, D.B., T.F. and C.R.; validation, D.B., T.F., K.J. and M.A.T.; formal analysis, D.B.; investigation, S.S., J.J.P., C.E.C., J.Š., D.H.G. and S.A.B.; resources, S.S., J.J.P. and C.E.C.; data curation, D.B. and C.M.; writing—original draft preparation, D.B.; writing, review and editing—All authors; supervision, D.B., M.A.T. and C.M.; project administration, D.B. and C.M.; funding acquisition, S.S. All authors have read and agreed to the published version of the manuscript.

**Funding:** This research was partially funded by Breakthrough Initiatives, Massachusetts Institute of Technology and Droplet Measurement Technologies.

**Institutional Review Board Statement:** Not applicable.

**Informed Consent Statement:** Not applicable.

**Data Availability Statement:** Not applicable.

**Acknowledgments:** We thank the extended Venus Life Finder Mission and the Rocket Lab engineering teams who have interfaced with the AFN engineering team to resolve issue related to integrating the AFN to the space capsule. The AFN design has a patent pending and is licensed by Droplet Measurement Technologies, LLC Cloud Measurement Solutions, LLC.

**Conflicts of Interest:** The authors declare no conflict of interest.